\documentclass[aps,prd,twocolumn,showpacs,nofootinbib,amsmath,amssymb,amsfonts,showkeys]{revtex4}

\usepackage{bm}
\usepackage{mathtext}
\usepackage[T2A]{fontenc}
\usepackage[T1,T2A]{fontenc}
\usepackage[utf8]{inputenc}
\usepackage[ukrainian,english]{babel}
\usepackage[dvips]{graphicx}
\usepackage{tikz}
\usepackage{graphicx,import}
\usepackage{color, colortbl}

\begin{document}

\title{Impact of the cosmological expansion on spectral energy density of radiation \\ in the intergalactic medium and once more about Olbers’ paradox}

\author{A. Yang$^1$, B. Novosyadlyj$^{1,2}$\footnote{Corresponding author}, G. Milinevsky$^{1,3,4}$\footnote{Corresponding author}}
\affiliation{$^1$International Center of Future Science and College of Physics of Jilin University, \\  
2699 Qianjin Str., 130012, Changchun, P.R.China, \\
$^2$Astronomical Observatory of Ivan Franko National University of Lviv, \\
8 Kyrylo and Methodij Str., 79005, Lviv, Ukraine,\\
$^3$Department of Atmosphere Physics and Geospace, National Antarctic Scientific Center,\\
16 Taras Shevchenko Blvd., Kyiv 01601, Ukraine
$^4$Main Astronomical Observatory of National Academy of Sciences of Ukraine, 27 Akademika Zabolotnoho Str., 03143, Kyiv, Ukraine}

\date{\today}
 
\begin{abstract}
We analyse the impact of the expansion of the Universe on the formation of the total spectral energy density of radiation in the intergalactic medium. Assuming the same proper thermal spectrum of sources, we show how the expansion of the Universe changes the nature of the energy distribution of the thermal spectrum: a decrease of the energy density in the Wien range and an increase in the Rayleigh--Jeans range with increasing the redshift of the bulk filled by sources. This is due to the cosmological redshift and the growing contribution of large number of distant sources. The numerical estimations  also illustrate the main factors that resolve the Olbers' paradox in the expanding Universe: i) the particle horizon, ii) the finiteness of the volume filled by luminous objects, and iii) the cosmological redshift. Applying the obtained expressions to the epoch of reionization made it possible to estimate the concentration of objects of various classes (stars, globular clusters, dwarf galaxies) necessary for complete reionization of hydrogen at $z=6$. It is shown that even a small part of globular clusters or dwarf galaxies with a thermal spectrum of moderate temperature, from those predicted by Press-Schechter formalism and its improvements, is able to completely reionize hydrogen in the intergalactic medium at $z=6$.
\end{abstract}
\maketitle

\section*{INTRODUCTION}

The studies of the spectral energy distribution of radiation in our galaxy Milky Way and other galaxies of different types are important problems of current astrophysics. Knowledge of such distributions makes it possible to estimate the ionization and thermal states of interstellar gas and dust, their luminescence and absorption in the lines and in the different ranges of electromagnetic radiation. We now know that the bulk of the mass density of the Universe resides in the hidden components: dark matter and dark energy. In terms of percentage, the first is about 25\%, the second is about 70\%. And only about 5\% is ordinary matter, which consists of stars, galaxies, interstellar gas and dust, the Sun, planets, the Earth and, ultimately, everything living and non-living that is on it. This type of matter in cosmology is called baryonic, after the name of the type (baryon) of the most massive particles - protons and neutrons, from which the nuclei of all known chemical elements are formed. The main part of baryonic matter, about 80\%, is in intergalactic medium. It is very rarefied, about one particle per cubic meter, and very hot, tens of millions of degrees. As for the great rarefaction, this is explained by the rate of expansion of the Universe and its age, 13.8 billion years. As for the temperature, not everything is clear here. Is such a high temperature of the intergalactic gas a consequence of the violent processes of energy release in the early Universe and the slow cooling of rarefied plasma, or is it supported by the radiation of modern galaxies of various types and quasars? To answer these questions, you need to know the spectral energy distribution of radiation in the expanding Universe. 

The luminous baryonic matter in the observed Universe is aggregated in the galaxies, which form the large scale structure of the Universe now called the Cosmic Web. Its elements are clusters of galaxies, superclusters, giant voids, sheets and filaments. In the paper, the following simplifications will be made in order to use the analytical and semi-analytical research methods: radiation sources will be considered as star-like, thermal, uniformly distributed in space with a given density. The appearance and  distribution of luminous matter could be crucial in the early stages of the evolution of the Universe, when the first stars and galaxies were formed. According to modern ideas, supported by observational data, the formation of the structure proceeded ``from the bottom up'', from stars to galaxies, clusters of galaxies and so on. In the young Universe, after a hot early stage and cosmological recombination, an epoch has come of almost neutral gaseous medium transparent to cosmic microwave background radiation (CMBR). At that time, there were no stars or galaxies. They have just begun to form from initial seeds generated in the early inflationary stage which can be considered as a component of the Big Bang itself. This epoch, called as Dark Ages, is now attracting great interest of researchers. For the observations of this era, the most advanced telescope of our time, the James Webb Space Telescope, was created.  

The aim of the paper is to study the impact of the expansion of the Universe on the formation of the average spectral energy density of radiation (SEDR) from the sources with a given space number density and blackbody spectrum. Such spectrum with some effective temperature can be approximation of energy spectrum of the first sources of radiation in the Cosmic Dawn epoch: stars, globular clusters and dwarf galaxies.  We apply this approach for study the possibility of complete reionization of intergalactic hydrogen by these sources, that is important in the context of ongoing discussions about their nature \cite{Bromm2011,Madau2015,Bouvens2015a,Robertson2015,Atek2015,Finkelstein2015,Bouvens2017,Dayal2018,Mason2019,Mistra2018,Ishigaki2018,Finkelstein2019,Naidu2020,Dayal2020,Robertson2022,Robertson2023,Bunker2023,Roberts2023,Mascia2023,Atek2024}. We apply it also for the numerical estimations of the total energy density from the extragalactic sources or glow of the night sky in the context  of ongoing discussions on the resolution of Olbers' paradox. Historical and current discussions of the last can be found in the papers \cite{Harrison1964,Belinfante1975,Wesson1986,PanGen1988,Harrison1990,Wesson1991,Maddox1991,Peebles1993,Couture2012,Conselice2016,Mattila2019,Harari2019} and in the references given in them. 

The outline of the paper is as follows. In the first section we develop the method of computation of averaged SEDR for static infinite eternal Universe and illustrate Olbers' paradox by estimation of the energy density of radiation from all stars. In the second section we study the mean SEDR from the thermal sources (stars) with different temperatures in the standard $\Lambda$CDM cosmological model of the Universe. Here we also resolve the Olbers' paradox in this approach. In the third one we use this method for study the SEDR in the early Universe when the first stars and galaxies appeared. We estimated the number density of thermal star-like sources at cosmological redshift 6 were intergalactic hydrogen becomes completely reionized. The last section contains the conclusions.

\section{Spectral energy distribution of radiation from star-like sources in the static infinite Universe: Olbers' paradox}

Let's suppose that the Universe is static, eternal in time and infinite in space. Further we call it as static infinite one. It is filled by stars with number density $n_*$ and radius $R_*$ each. Suppose also that each star is thermal source of radiation with temperature $T_*$. Let's estimate the SEDR at an arbitrary point, which is equidistant from the nearest stars. We consider the case of empty space between stars for beginning. 
 
The solid angle of star visible from distant $r$ is $\Omega_*=\pi R_*^2/r^2$. The energy density of its radiation there at frequency $\nu$ per unit range of $\nu$ is as follows
\begin{equation}
\epsilon_\nu=\frac{\pi R_*^2}{r^2}\frac{1}{c}B_\nu ,\label{epsnu1}
\end{equation}
where $B_\nu$ is Planck function,
\begin{equation}
B_{\nu }(\nu ,T_*)=\frac{2h\nu^{3}}{c^{2}}\frac {1}{e^{h\nu/kT_*}-1}  \quad    \left[\mathrm{\frac{erg}{s\cdot cm^2\cdot Hz\cdot sr}}\right]. \label{Bnu}
\end{equation}
Here, $T_*$ is the temperature of star photosphere in Kelvins, $h$ and $k$ are Planck and Boltzmann constants accordingly, and $c$ is the speed of light in the vacuum. 
We neglect here the effect of the darkening of the star's disk towards the edge and shielding the distant stars by closer ones.

The number of stars in the spherical shell of radius $r$ and thickness $dr$ ($\gg R_*$) is $dN=4\pi r^2drn_*$. The SEDR from all stars in such shell is $dE_\nu=\epsilon_\nu dN$. The total SEDR at frequency $\nu$ from all stars in spherical volume with radius $r_u$ is integral over $r$,
\begin{equation}
E_\nu=\int_0^{r_u}\frac{\pi R_*^2}{r^2}\frac{1}{c}B_\nu\cdot n_*4\pi r^2dr=\frac{4\pi^2 R_*^2n_*}{c}B_\nu r_u
\label{enu1}
\end{equation}
in units erg/s$\cdot$cm$^3\cdot$Hz. 
The total energy density is integral of SEDR over $\nu$ from 0 to $\infty$ which gives us 
\begin{eqnarray*}
E=\int_0^\infty E_\nu d\nu=\frac{4\pi R_*^2n_* r_u}{c}\sigma T_*^4,
\label{e1}
\end{eqnarray*}
where $\sigma$ is Stefan-Boltzmann constant.

One can see that $E_\nu$ and $E$ $\rightarrow\infty$, when $r_u\rightarrow\infty$. So, in the infinite eternal Universe filled by point-like sources of light the SEDR and total energy of radiation is infinite. If we take into account the size of stars the whole sky will be covered with discs of stars in the volume with some finite $r_*$. It means that each patch of sky will glow like the photosphere of the star. But real nights are dark. It is classical photometric paradox which has been described in 1823 and reformulated in 1826 by Heinrich Olbers.

It may seems that this paradox is resolved if there is absorbing interstellar gas. Let's consider now the static infinite Universe filled homogeneously by stars in absorption interstellar medium. Such medium decreases the intensity of each source by factor $e^{-\tau}$, where $\tau\equiv\kappa_\nu r$ is optical thick between observer and source of radiation. So, the SEDR from it at frequency $\nu$ is 
$$\epsilon_\nu=\frac{\pi R_*^2}{r^2}\frac{1}{c}B_\nu e^{-\kappa_\nu r}.$$
It is exact solution of the radiative transfer equation $dI_\nu/dr=-\kappa_\nu I_\nu + \eta_\nu$ in the medium with negligible emissivity. Here $I_\nu$ is intensity at frequency $\nu$, $\kappa_\nu$ is absorption coefficient, $\eta_\nu$ is emission coefficient of unit volume of medium at frequency $\nu$. Repeating the speculations from the case of empty space we obtain the SEDR at frequency $\nu$ from all stars in spherical volume with radius $r_u$
\begin{equation}
E_\nu =\frac{4\pi^2 R_*^2n_*}{c}B_\nu\frac{1-e^{-\kappa_\nu r_u}}{\kappa_\nu}.
\label{enu2}
\end{equation}
First of all, let's make sure that this equation is more general than (\ref{enu1}). Really, $\frac{1-e^{-\kappa_\nu r_u}}{\kappa_\nu}\rightarrow r_u$ when $\kappa_\nu\rightarrow 0$, and (\ref{enu2}) $\rightarrow$ (\ref{enu1}).

 \begin{figure*}[htb!]
\includegraphics[width=0.48\textwidth]{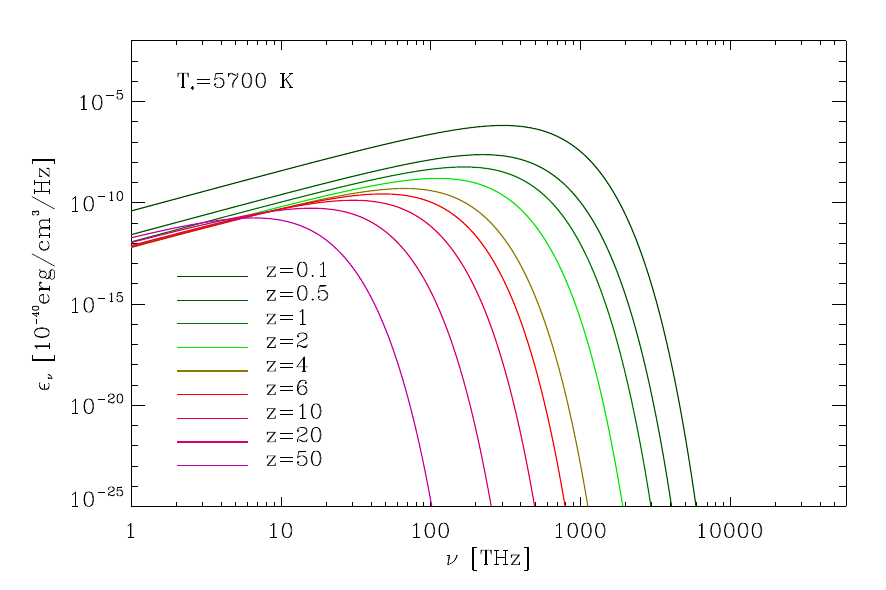}
\includegraphics[width=0.48\textwidth]{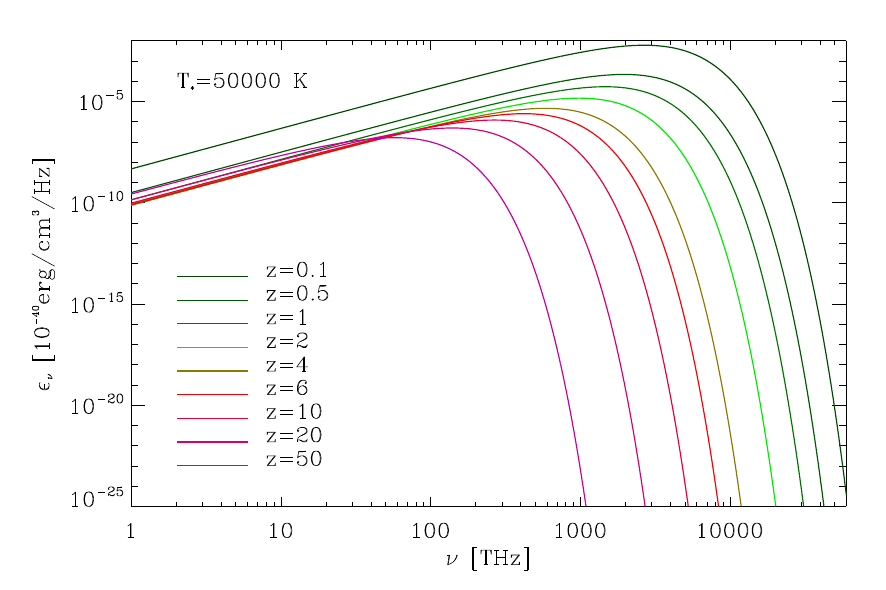}
\caption{The SEDR at z = 0 from individual stars with temperatures of 5700 K (left) and 50000 K (right) located at different redshifts z.}
\label{fig1}
\end{figure*}

Now, let's use (\ref{enu2}) for infinite eternal Universe ($e^{-\kappa_\nu r_u}\rightarrow0$ when $r_u\rightarrow\infty$): 
\begin{equation}
E_\nu=\frac{4\pi^2 R_*^2n_*}{c\kappa_\nu}B_\nu, \quad E=\frac{4\pi^2 R_*^2n_*}{c\kappa}\sigma T_*^4.
\label{enu3}
\end{equation}
Both value, SEDR and total energy density, are finite for infinite eternal Universe.  
Thus (\ref{enu3}) looks as solution of Olbers' paradox for infinite eternal Universe. But it is not so. To show this we compare $E$ with some realistic values of $n_*$ and $\kappa_\nu$ in (\ref{enu3}) with solar $\epsilon_\odot$ near the Earth. 
We use the value of optical depth $\tau=0.0544$ to the reionization epoch at $z_{rei}=7.67$ \cite{Planck2020} which gives the estimation of $\kappa\approx6\cdot10^{-9}$ kpc$^{-1}$. 
The comoving number density of stars $n_*$ we estimate in the following way: we divide the number of stars in our Galaxy ($2\cdot10^{11}$) by spherical volume with galaxy radius (15 kpc) and  divide this value by 200, density contrast of virialised halo. Thus, we obtain $n_*^{max}\approx7\cdot10^4$ kpc$^{-3}$. It is very rough estimation but it is not critical and below results can be easy renormalized to more realistic value, which is unknown, unfortunately. It is upper limit for $n_*$. 
The lower limit can be estimated from the number density stars in our Galaxy and the estimated number of such galaxies $\sim10^{13}$ in the observable Universe up to $z_{rei}$ (comoving distance $r(z_{rei})\approx9\cdot10^6$ kpc). In this case we obtain $n_*^{min}\approx700$ kpc$^{-3}$. 
Using eqs. (\ref{epsnu1}) and (\ref{enu3}) and assuming $T_*\sim T_\odot$ we obtain that $E/\epsilon_\odot=4\pi n_*r_\odot^2T_*^4/\kappa/T_\odot^4\sim4\cdot10^{-5}-4\cdot10^{-3}$. This means that the brightness of the night sky for the range $n_*^{min}-n_*^{max}$ is between full Moon night brightness to twilight one during sunrise/sunset, that is thousand times larger than brightness of starlight in moonless night sky. Another factor not taken into account will increase the brightness of night sky: the heating of absorbing medium by radiation leads to the thermodynamic equilibrium between them. In the infinite eternal Universe the interstellar medium will shine like surface of stars. Therefore Olbers' paradox has not solution in the infinite eternal Universe even with interstellar absorption medium. 
 
\section{Spectral energy distribution of radiation from star-like sources in the expanding Universe}

Let's estimate now the energy density of radiation in the expanding Universe, which filled homogeneously by the star-like sources with radius $R_*$, temperature $T_*$ and comoving number density $n_*$. We also accept that the interstellar medium is optically thin, global 3-space is Euclidean and the main parameters of standard $\Lambda CDM$ cosmological model are as follows: the Hubble constant is $H_0=67.36$ km/s/Mpc, dimensionless matter density parameter is $\Omega_m=0.3153$, dimensionless cosmological constant is $\Omega_\Lambda=0.6847$ and mass density fraction of primordial helium $Y_p\equiv \rho_{He}/(\rho_H+\rho_{He})=0.2446$ \cite{Planck2020}. 
 
The rate of expansion of the Universe is described by the first Friedmann equation
$$H(z)\equiv \frac{1}{a}\frac{da}{dt}=H_0\sqrt{\Omega_m (z+1)^3 + \Omega_\Lambda},$$
where $a(t)$ is scale factor, which describes the expansion of the Universe and $z$ is cosmological redshift. Both values $a$ and $z$ are connected by simple relation $z=a^{-1}-1$, which also means that for current epoch $t_0$  $a(t_0)=1$. In the expanding Universe, the  emitted or proper frequency $\nu_e$ at $z_e$ related with observed frequency $\nu_o$ at $z_o$ via cosmological redshifts as follows
\begin{equation}
 \nu_o=\nu_e\frac{1+z_o}{1+z_e}. \label{nuz}
\end{equation} 
\begin{figure*}[htb!]
\includegraphics[width=0.48\textwidth]{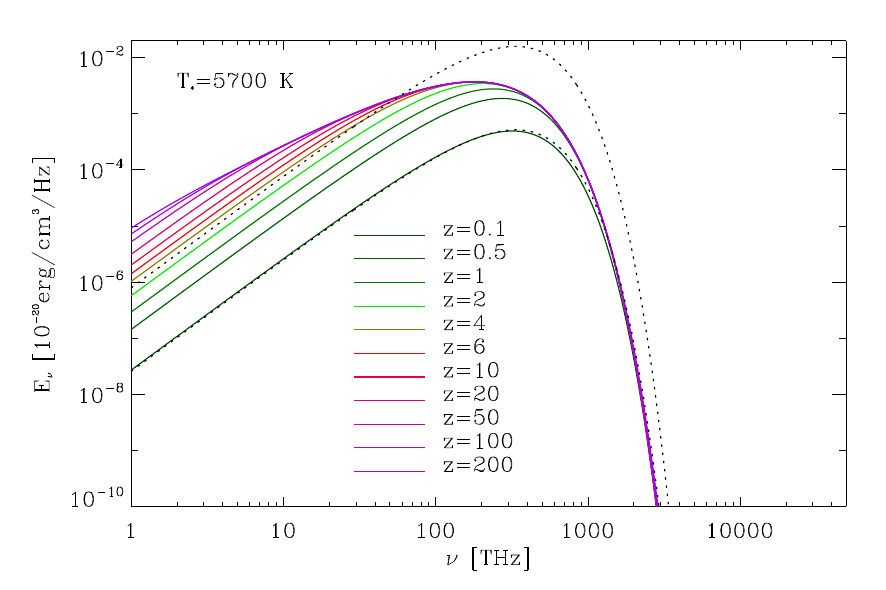}
\includegraphics[width=0.48\textwidth]{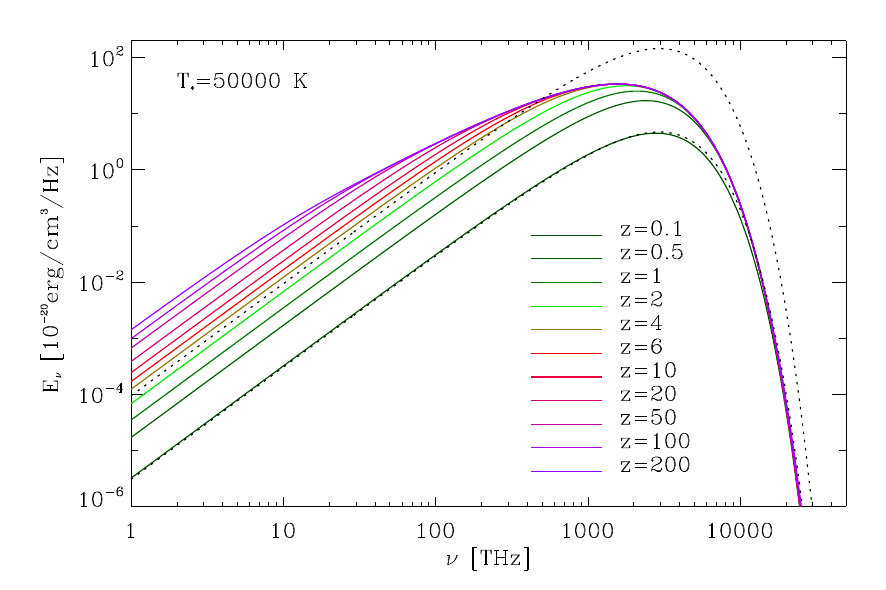}
\caption{The SEDR at $z=0$ from stars with temperature 5700 K (left) and 50000 K (right) which are in the volume with comoving radius $r_i(z_i)$ ($z_i$=0.1, 0.5, 1, 2, 4, 6, 10, 20, 50, 100, 200) in the standard $\Lambda$CDM cosmological model. Lower and upper dotted lines show the spectral energy density of blackbody radiation in the static infinite flat Universe from the stars  with corresponding temperatures in the volumes with radiuses corresponding to the comoving distance at $z=0.1$ (lower) and $z=200$ (upper) of expanding Universe.} \label{fig2}
\end{figure*}
\begin{figure*}[htb!]
\includegraphics[width=0.48\textwidth]{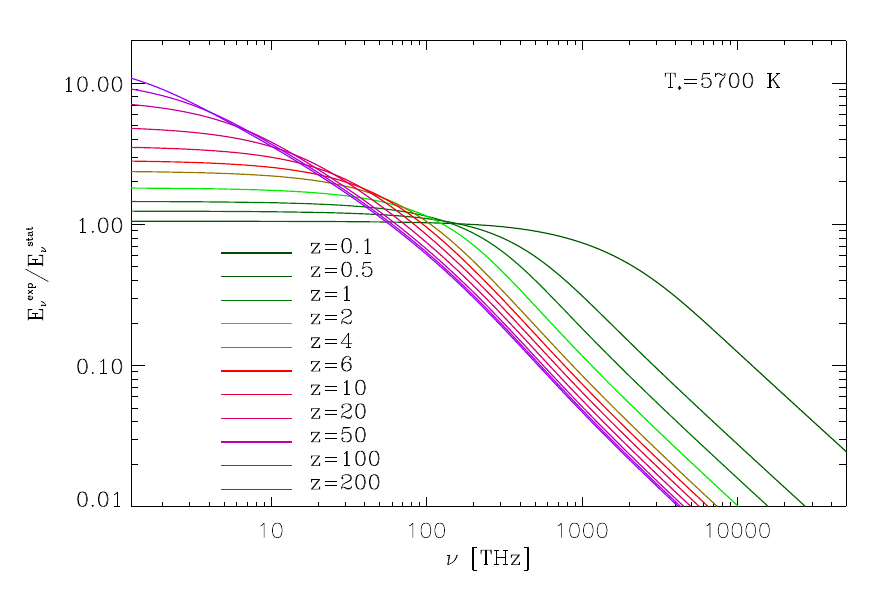}
\includegraphics[width=0.48\textwidth]{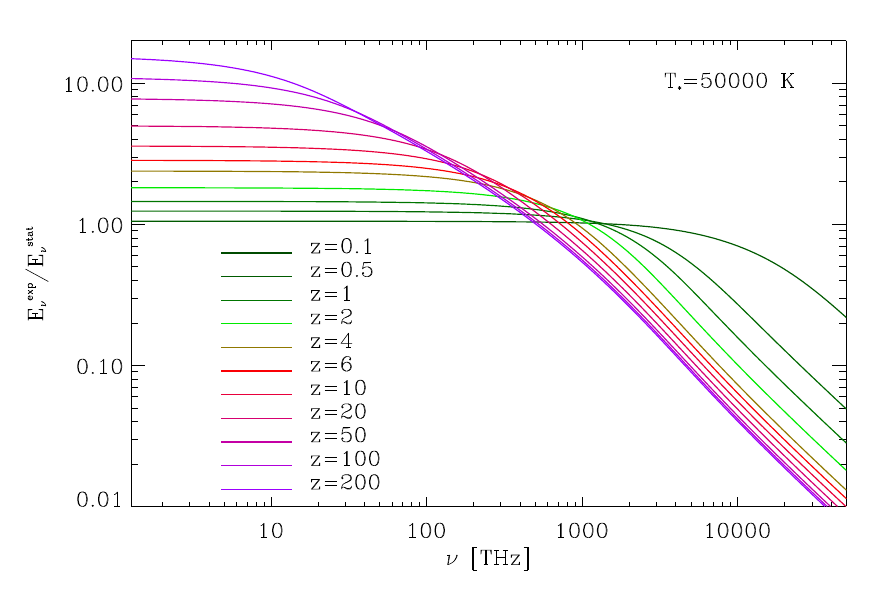}
\caption{Ratios of SEDR from stars which are in the volume with comoving radius $r_i(z_i)$ ($z_i$=0.1, 0.5, 1, 2, 4, 6, 10, 20, 50, 100, 200) in the expansion Universe to SEDR of the same stars in the same volumes in static infinite flat Universe.}\label{fig3}
\end{figure*}

The intensity of radiation at frequency $\nu$ emitted by the unit surface of star, which is at redshift $z$, in the rest frame of star is described by Planck function (\ref{Bnu}).
The proper spectral luminosity at frequency $\nu$ in the range $\nu,\,\nu+d\nu$ there is as follows
\begin{equation}
L_\nu(z)d\nu=4\pi R_*^2\pi B_{\nu}d\nu\quad \left[\mathrm{\frac{erg}{s}}\right].\label{Lz}
\end{equation}
The observed luminosity at $z=0$ is related with luminosity at any $z>0$ as follows
\begin{equation}
L_{\nu'}(0)d\nu'= \frac{L_\nu(z)d\nu}{(1+z)^2}.
\label{L0Lz}
\end{equation}
where observed and emitted frequencies are related by redshift $z$, $\nu'=\nu/(1+z)$. Expression (\ref{L0Lz}) can be easily understood: if star emits N quanta per second, each with energy $h\nu$, then the time intervals at star and at observer are related as $dt(0)=(1+z)dt(z)$. So, the observer during his own (proper) one second detects $N/(1+z)$ quanta, each with energy $h\nu/(1+z)$. 
 
Combining of equations (\ref{Lz}) and (\ref{L0Lz}) gives
\begin{equation}
L_{\nu'}(0)d\nu'=\frac{4\pi R_*^2\pi B_{\nu}d\nu}{(1+z)^2} \quad \left[\mathrm{\frac{erg}{s}}\right].
\end{equation}
\begin{figure*}[htb!]
\includegraphics[width=0.48\textwidth]{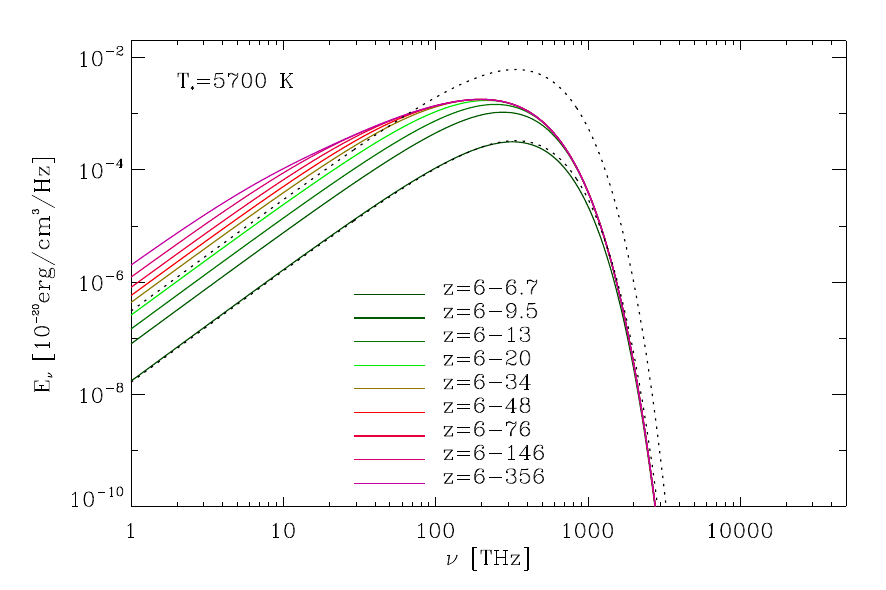}
\includegraphics[width=0.48\textwidth]{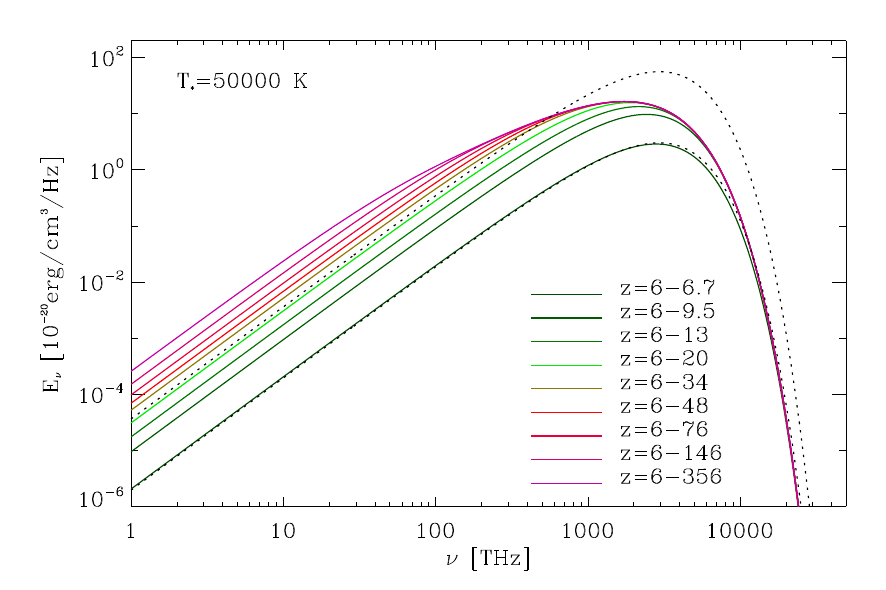}
\caption{The SEDR at $z=6$ from stars with temperature 5700 K (left) and 50000 K (right) which are in the volume with comoving radius $r_i(z_i)-r(6)$ in the standard $\Lambda$CDM cosmological model. Lower and upper dotted lines show the spectral energy density of blackbody radiation in the static infinite flat Universe from the stars with corresponding temperatures in the volumes with radiuses corresponding to comoving distance from $z=6$ to $z=6.7$ (lower) and from $z=6$ to $z=356$ (upper) of expanding Universe.}\label{fig4}
\end{figure*}
\begin{figure*}[htb!]
\includegraphics[width=0.48\textwidth]{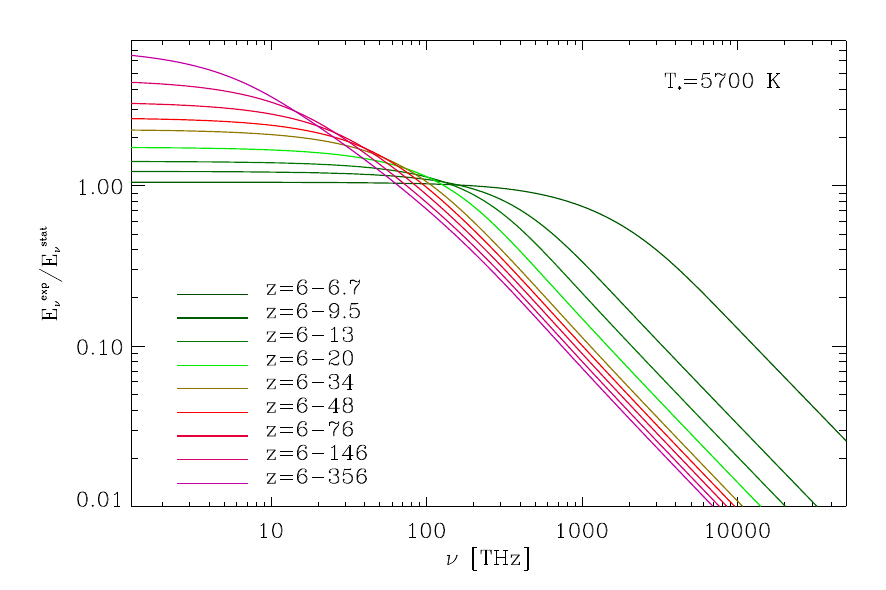}
\includegraphics[width=0.48\textwidth]{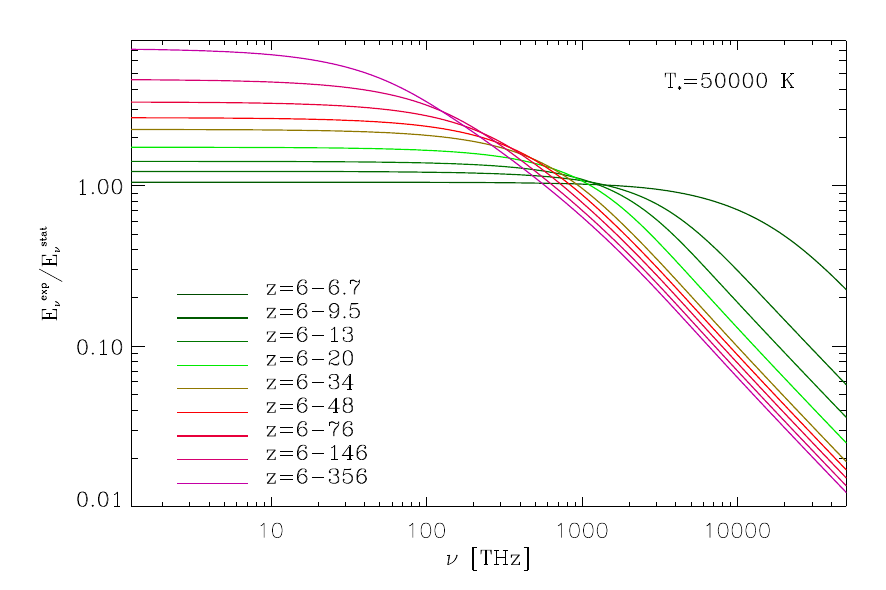}
\caption{Ratios of SEDR from stars which are in the volume with comoving radius $r_i(z_i)-r(6)$ in the expansion Universe to SEDR of the same stars in the same volumes in static infinite flat Universe.}\label{fig5}
\end{figure*}
Then the spectral flux\footnote{It is defined at hypersurface $t_0=const$, where $a=a_0=1$ and square of sphere with comoving radius $r$ is equals $S=4\pi r^2$.}  at frequency range $\nu,\,\nu+d\nu$ and SEDR from single star at redshift $z$ in the flat Universe ($K=0,\,\chi(r)=r$) are as follows
\begin{eqnarray} 
F_{\nu'}(0)d\nu'&=&\frac{\pi R_*^2}{(1+z)^2r^2} B_{\nu}d\nu \quad \left[\mathrm{\frac{erg}{cm^2\,s}}\right],\label{Fnu'} \\
\epsilon_{\nu'}(0)d\nu'&=&\frac{\pi R_*^2}{(1+z)^2r^2} \frac{1}{c}B_{\nu}d\nu \quad \left[\mathrm{\frac{erg}{cm^3}}\right]. \label{epsnu'}
\end{eqnarray}
If we use (\ref{Bnu}) and substitute $\nu$ by $\nu'(1+z)$ and $d\nu=(1+z)d\nu'$, and omit a prime near $\nu$ in the left and right hand sides then the  expression for SEDR (\ref{epsnu'}) becomes
\begin{equation}
 \epsilon_{\nu}(0)d\nu=\frac{\pi R_*^2}{c^3r^2}\frac{2h\nu^{3}(1+z)^2}{e^{h\nu(1+z)/kT_*}-1}d\nu \quad \left[\mathrm{\frac{erg}{cm^3}}\right],\label{eps0}
\end{equation}
where comoving distance can be computed for given $z$ as
\begin{equation}
r=\frac{c}{H_{0}} \int^{z}_{0} \frac{dz'}{\sqrt{\Omega_{m}(z'+1)^{3}+\Omega_{\Lambda}}}.
\label{rz0}
\end{equation}

Taking into account the relation (\ref{nuz}), the equations (\ref{L0Lz})-(\ref{rz0}) can be easily generalized for computations of SEDR at any $z_o>0$. For that we throughout substitute factor $(1+z)$ by $(1+z)/(1+z_o)$ in (\ref{eps0}), and the lower limit in the integral (\ref{rz0}) by $z_o$. This is true as long as we do not take into account the evolution of luminosities of sources.

In Fig. \ref{fig1}, we show the SEDR at $z=0$ from a single star with  $T_*=5700$ K and 50000 K, which is at different $z$. The amplitude of SEDR from a single star decreases with increasing $z$ over the entire frequency range, remaining blackbody with temperature $T_*/(1+z)$. The peak frequency of SEDR, $\nu_{peak}$, defined as Wien displacement law, decreases $\propto(1+z_o)/(1+z)$. The hydrogen atoms can be ionized by photons with energy $>13.6$ eV, that corresponds $\nu>3300$ THz. The SEDR's in Fig. \ref{fig1} show, that in the expanding Universe the solar-like stars (left panel) from close vicinity only, $z\lesssim2z_o+1$, can ionize the interstellar hydrogen atoms. For $z_o=0$, $z<1$, for $z_o=6$ (epoch of complete reionization) $z<13$, and space number density of such stars must be very large. Contrary, the SEDR's of hot stars from essentially large redshifts ($z\lesssim10z_o+9$ for case in the right panel) contain the large number of ionizing quanta. Such stars can reionize the intergalactic medium at $z\sim6$ with essentially lower space number density. In the next section we estimate these values.  

Now, we estimate the SEDR from all stars homogeneously filled in the Universe with comoving number density $n_*$. The number of stars in the spherical shell of comoving radius $r$ and thick $dr$ is $dN=4\pi n_*r^2dr$. The SEDR from all stars in such shell is $dE_\nu(0) d\nu=\epsilon_\nu(0) d\nu dN$, where $\epsilon_\nu d\nu$ is (\ref{eps0}). The total SEDR at frequency $\nu$ in the range $d\nu$ from all stars in spherical volume with radius $r_u$ is obtained by integration over $r$ from  0 to $r_u$:
\begin{equation}
E_\nu d\nu=\int_0^{r_u} \frac{\pi R_*^2}{c^3r^2}\frac{2h\nu^{3}(1+z)^2}{e^{h\nu(1+z)/kT_*}-1}  d\nu\cdot n_*4\pi r^2dr, \nonumber
\end{equation} 
where $r_u$ is computed for given $z$. In the expanding Universe with $\Omega_m\neq0$ the maximal value of redshift is not limited, $max{\,z}\rightarrow\infty$, but $r_u$ is limited by particle horison, and $E_\nu d\nu$ is finite too. Taking into account (\ref{rz0}) we obtain finally
\begin{eqnarray}
E_\nu(0,z) d\nu&=&\frac{8\pi^2 h\nu^3n_*R_*^2}{c^2H_0}\label{Enu0} \\ 
&&\hskip-2cm\times\left[\int_0^z\frac{(1+z')^2}{e^{h\nu(1+z')/kT_*}-1}\frac{dz'}{\sqrt{\Omega_{m}(z'+1)^{3}+\Omega_{\Lambda}}} \right]d\nu. \nonumber
\end{eqnarray} 
To obtain SEDR at any $z_o$, we must to put the lower limit of the integral $z_o$, and substitute $(1+z')$ by $(1+z')/(1+z_o)$:
\begin{eqnarray}
E_\nu(z_o,z) d\nu&=&\frac{8\pi^2 h\nu^3n_*R_*^2}{c^2H_0}\label{Enuz0}\\ 
&&\hskip-2.5cm\times\left[\int_{z_o}^z\frac{[(1+z')/(1+z_o)]^2}{e^{h\nu(1+z')/kT_*(1+z_o)}-1}\frac{dz'}{\sqrt{\Omega_{m}(z'+1)^{3}+\Omega_{\Lambda}}} \right]d\nu. \nonumber
\end{eqnarray} 
 
The results of computations of SEDR at $z_o=0$ from stars with temperature 5700 K (left) and 50000 K (right) which are in the volume with comoving radius $r_i(z_i)$ ($z_i$=0.1, 0.5, 1, 2, 4, 6, 10, 20, 50, 100 and 200) in the standard $\Lambda$CDM are presented in Fig. \ref{fig2}. One can see how the expansion of the Universe changes the energy distribution of composite SEDR formed by sources with the same proper blackbody spectrum. The dotted lines show the undistorted spectrum in a static infinite model of the Universe with the same number density of the same stars. The ratios of spectral densities of radiation in the expanding and static infinite cosmological models are shown in Fig. \ref{fig3}. The compaction of lines with increasing $z$ indicates the existence of an enveloping asymptote at $z\gtrsim 200$, existence of which follows from the expression (\ref{Enu0}). Analogical results for $z_o=6$ for the same values of $(1+z)/(1+z_o)$ as for $z_o=0$ are shown in Fig. \ref{fig4}-\ref{fig5}\footnote{According to the standard cosmological model the first luminous sources appeared at $z\lesssim 30-50$. In the toy model here the sources at huge z are for illustration of effect only}.

\begin{figure}[htb!]
\includegraphics[width=0.48\textwidth]{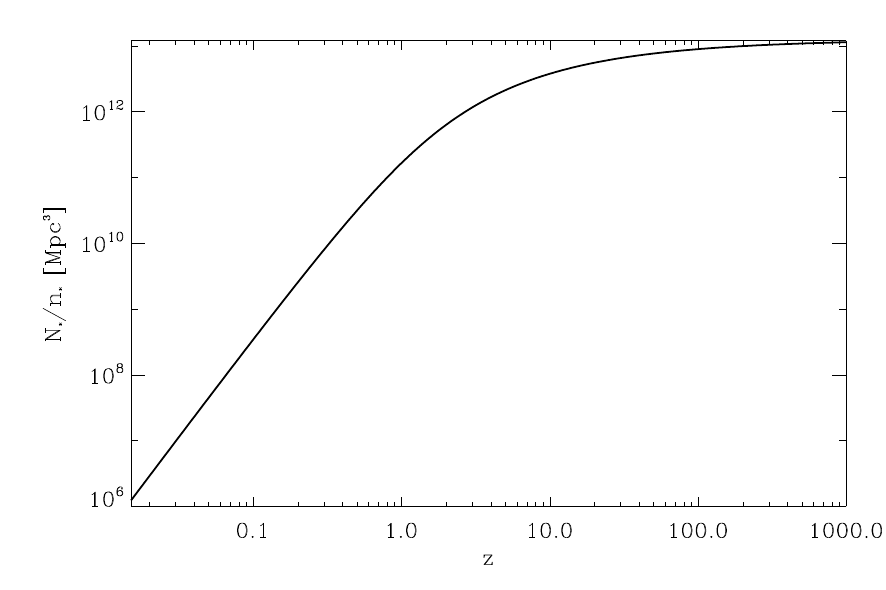}
\caption{The dependence of number of sources on redshift of edge of volume filled by sources in the expanding Universe. }
\label{fig6}
\end{figure}

The SEDR's shown in Figs. \ref{fig2} and \ref{fig4} are caused by two reasons: the cosmological redshift (\ref{nuz}) and the dependence of number of sources on redshift of edge of bulk filled by sources in the expanding Universe, $N_*(z)$. The last dependence in the case $n_*=const$ at $z'\le z$ and $n_*=0$ at $z'>z$ is as follows
\begin{equation}
N_*(z)=\frac{4}{3}\pi n_*\left[\frac{c}{H_0}\int_0^{z}\frac{dz'}{\sqrt{\Omega_m(1+z')^3+\Omega_\Lambda}}\right]^3. \nonumber
\end{equation}
The ratio $N_*(z)/n_*$ is shown in Fig. \ref{fig6}. The finiteness of number of sources in the Big Bang Universe is partial resolution of Olbers' paradox.
The evolution effects and variety of sources, which are not taken here into account, are important for the computation of SEDR of course. They can be taken into account correctly in the complete cosmological simulations only.

For estimations, the radius of the stars with $T_*=T_\odot=5700$ K we assume to be equal to the radius of the Sun $R_*=R_\odot=6.96\cdot10^{10}$ cm. For massive stars with high temperature we use the well known relations $T/T_\odot\approx\left(M/M_\odot\right)^{0.6}$, $R/R_\odot\approx2.67\left(M/M_\odot\right)^{0.36}$. They lead to relation $R_*\approx2.67\left(T_*/T_\odot\right)^{0.6}R_\odot$, which for $T_*=50000$ K gives $R_*\approx6.84\cdot10^{11}$ cm. The comoving number density of stars $n_*$ we  assumed to be equal $n_*^{max}$, estimated in the previous section. The product $n_*R_*^2$ is a normalisation constant in eqs. (\ref{Enu0})-(\ref{Enuz0}). 

Let's estimate the maximal square of the sky covered by stars in the volume from $z=0$ up to 100. The angular area of star, which is at the comoving distance $r(z)$, equals $\omega=\pi (1+z)^2 R_*^2/r^2$. So, all stars in this volume cover
\begin{equation}
 \frac{\Omega}{4\pi}=\frac{\pi c n_*^{max}R_*^2}{H_0}\int_0^{100}\frac{(1+z)^2 dz}{\sqrt{\Omega_m (1+z)^3+\Omega_\Lambda}} \nonumber
\end{equation}
part of the sky.  For sun-like stars it is about $10^{-6}$, for brighter ones analysed here, it is about $10^{-4}$. Therefore, only tiny part of the sky is covered by stars in the Big Bang model of the Universe, that resolve the classical Olbers' paradox.

The cosmological redshift changes the spectral energy distribution from high frequencies to low ones. To visualise this we added the dotted lines which show the spectral energy density of blackbody radiation in the static infinite flat Universe from the stars with the same temperatures in the same volumes defined by $z=0.1$ (lower) and $z=200$ (upper) for case $z_o=0$ (Fig. \ref{fig2}) and by $z=6.7$ (lower) and $z=356$ (upper) for case $z_o=6$ (Fig. \ref{fig4}) in the expanding Universe. The lines in Figs. \ref{fig4}-\ref{fig5} are close to corresponding lines in Figs. \ref{fig2}-\ref{fig3}. This is because the calculations are made for the same values $(1+z_i)/(1+z_o)$ for both cases $z_o=0$ and $z_o=6$. The small differences are caused by those that for $z_o=0$ the cosmological model is $\Lambda$CDM, and at $z_o=6$ the cosmological model is closer to Einstein-de Sitter one.

Since, the thermal spectrum is continuous the energy density of radiation at an arbitrary frequency is the contribution of stars from the entire bulk in the sphere with radius $r_i(z_i)$. The ratio $E^{exp}_\nu/E^{stat}_\nu$ is $<1$ in the Wien range and $>1$ in the Rayleigh-Jeans one, and monotonic increases from high frequency to low one and grows faster for larger bulks of stars. The peak frequency of SEDR, $\nu_{peak}$, slowly decreases with increasing $(1+z)/(1+z_o)$, and practically freezes for $(1+z)/(1+z_o)>8$ at $\nu_{peak}\approx180$ THz for stars with 5700 K, and at $\approx1600$ THz for stars with 50000 K. The ratio $E^{exp}_\nu/E^{stat}_\nu$ becomes $>1$ at lower frequencies, $\sim0.3\div0.6\,\nu_{peak}$, for both type of stars. Figures \ref{fig2}-\ref{fig3} and \ref{fig4}-\ref{fig5} illustrate that well. 

We must also to note that integrals in (\ref{Enu0})-(\ref{Enuz0}) asymptotically approach the finite small value when $(1+z)/(1+z_o)\rightarrow\infty$. Really, its part in the redshift range $(1+z)/(1+z_o)\ge100$ for any fixed $\nu$ can be reduced to the simple integral $I_z=C\int_{a}^\infty\sqrt{x}e^{-x}dx$, where $x=h\nu(1+z)/kT_*(1+z_o)$, $a=100h\nu/kT_*$ and $C=(kT_*/h\nu)^{3/2}/\sqrt{\Omega_m(1+z_o)}$, which approximately equals $C\sqrt{a}/e^{-a}$. Its numerical value is lower for higher $\nu$ and lower $T_*$, that is illustrated by lines for highest $z$ in Fig. \ref{fig2}-\ref{fig5}.

Let's estimate the total energy density of radiation from all stars which exist from $z=30$, for example, and compare it with the energy density of solar radiation. To do that we must to integrate the expression (\ref{Enuz0}) over $\nu$ from 0 to $\infty$: 
\begin{equation}
E_{all\,stars}= \int_0^\infty E_\nu(0,30)d\nu. \nonumber
\end{equation} 
The numerical integration gives: $E_{all\,stars}\approx3\cdot10^{-21}$ erg/cm$^3$ for stars with 5700 K and $E_{all\,stars}\approx1\cdot10^{-18}$ erg/cm$^3$ for stars with 50000 K. They are essentially lower then energy density of cosmic background radiation $E_{CMB}\approx6\cdot10^{-13}$ erg/cm$^3$. The total flux from Sun at the Earth is $F_{Sun}=1.367\cdot10^6$ erg/s/cm$^2$. It is called the solar constant. The energy density of solar radiation $E_{Sun}=F_{Sun}/c=4.56\cdot10^{-5}$ erg/cm$^3$. So, the total energy density of radiation from all stars with temperature 5700 K (50000 K) at the night side of the Earth is by $\approx 2\cdot10^{16}$ ($\approx 5\cdot10^{13}$) times lower than the total energy density of radiation from the Sun at the day side of the Earth. 

It is the resolution of Olbers' paradox ''Why the night sky is dark?''. The results, presented above, illustrate that the main reasons are: 1) a particle horizon in the Big Bang model (spatial finiteness of observable space), 2) spatial finiteness of volume filled by stars, 3) cosmological redshift. The interesting historical discussions about resolutions of this paradox can be found in a lot of papers and books, in particular \cite{Harrison1964,Belinfante1975,Wesson1986,PanGen1988,Harrison1990,Wesson1991,Maddox1991,Peebles1993,Couture2012,Conselice2016,Mattila2019,Harari2019}. We have presented here an original, mathematically simple resolution of Olbers' paradox in the standard cosmological model.

\section{On the possibility of cosmological reionization by the first stars, globular clusters and dwarf galaxies}

\begin{figure}[htb!]
\includegraphics[width=0.48\textwidth]{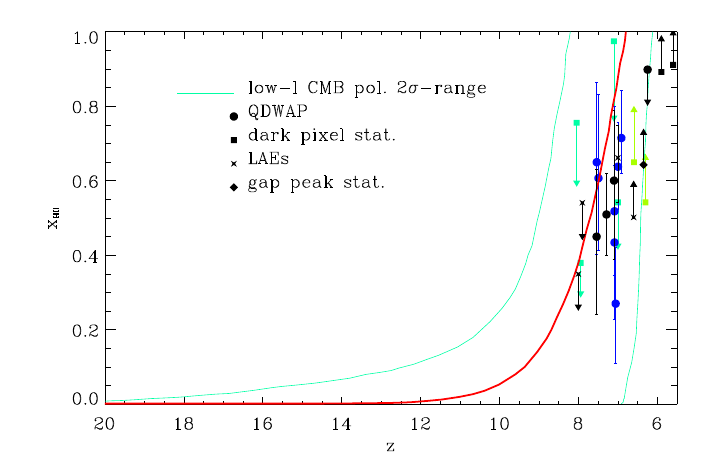}
\caption{Fraction of ionized hydrogen at Cosmic Dawn and Reionization epochs from cosmological and astrophysical observational data. The green thin solid lines show the 2$\sigma$-range given by Planck Collaboration \citep{Planck2020}, the red thick solid line is the median value \citep{Glazer2018}, the circles show the estimations derived from the damping wing absorption profiles in the spectra of quasars (QDWAP)  \cite{Schroeder2013,Greig2017,Mortlock2011,Davies2018,Bouvens2015,Banados2018,Greig2022}, the squares -- from the dark pixel statistics \cite{McGreer2015}, 4-fold stars -- from the redshift-dependent prevalence of $Ly\alpha$ emitters (LAEs) \cite{Schenker2014,Mason2018,Mason2019a,Ouchi2010}, the diamonds -- from the gap/peak statistics \cite{Gallerani2008}.} \label{fig7}
\end{figure}

The number density of the first sources of UV radiation increases at the end of the first billion years after the Big Bang, the degree of ionisation of hydrogen increases, and the epoch of Reionization begins. A few types of observational data independently support such increasing of fraction  of ionized hydrogen. They are shown in Fig. \ref{fig7}. It was shown \cite{Novosyadlyj2022,Novosyadlyj2023,Novosyadlyj2024} that intergalactic radiation with the composite thermal spectral energy distributions satisfy the observational constraints on the dependence of ionized fraction of the hydrogen on redshift at the Cosmic Dawn and Reionization epochs. The position, depth and profile of the absorption 21 cm hydrogen line, redshifted to the meter wavelength by cosmological expansion, can be observational test for SEDR at these epochs \cite{Novosyadlyj2023,Novosyadlyj2024}.

Let's assume that number of photoionization by ultraviolet radiation of stars and recombination per second in unit of volume are equal:
\begin{equation}
n_{HI}(z) R_{1c}(z)=\alpha_r(T) n_e(z) n_{HII}(z),\nonumber
\end{equation}
where $\alpha_r$ is the total rate of recombination on all levels, $n_e$ is number density of free electron, $n_{HII}$ is number density of ionised hydrogen, $n_{HI}$ is number density of neutral hydrogen and $R_{1c}$ is the rate of photoionization of atomic hydrogen from the first level by radiation of the stars. It is useful to introduce the fraction of ionized hydrogen atoms $x_{HII}\equiv n_{HII}/n_H$, the fraction of neutral hydrogen atoms $x_{HI}=n_{HI}/n_H$, where $n_H=n_{HI}+n_{HII}$ is total number density of hydrogen nuclei. At this epoch ($z=6$) only hydrogen is ionized, helium atoms and other chemical elements are neutral. So, $n_e=n_{HII}=x_{HII}n_H$. The equation of ionisation-recombination equilibrium with such notations becomes
\begin{equation}
(1-x_{HII})R_{1c}(z)=\alpha_r(T) x_{HII}^2n_H(z),\label{ire}
\end{equation}

The total number density of hydrogen nuclei can be estimated using the values of cosmological parameters in the standard $\Lambda$CDM (presented at the beginning of section II). The total mass-energy density (baryonic matter + dark matter + dark energy) at current epoch is as follows: $\rho_{cr}^0=3H_0^2/8\pi G$, where $G$ is Newtonian gravitational constant.
The mass density of hydrogen and number density of hydrogen atoms at current epoch are 
$\rho_H^0\approx\Omega_b\rho_{cr}(1-Y_p)$ and $n_H^0\approx\rho_H^0/m_H$, 
where $m_H$ is mass of hydrogen atom and $Y_p$ is primordial mass fraction of helium. The number density of hydrogen atoms at any redshift $z$ is $n_H(z)=n_H^0(1+z)^3$.
So, the general expression for $n_H(z)$ is as follows
\begin{equation}
n_H(z)=\frac{3H_0^2\Omega_b(1-Y_p)}{8\pi Gm_H}(1+z)^3.\nonumber
\end{equation}
At $z=6$ it is $n_H(z=6)\approx7\cdot10^{-5}$ cm$^{-3}$.

We use the analytical approximation of the total rate of recombination on all levels proposed by \cite{Verner1996}:
\begin{equation}
\alpha_r(T)=\frac{a}{\sqrt{T/T_1}\left(1+\sqrt{T/T_1}\right)^{1-b}\left(1+\sqrt{T/T_2}\right)^{1+b}}\nonumber,
\end{equation}
with the coefficients $a=7.982\cdot10^{-11}$, $b=0.748$, $T_1=3.148$ K, $T_2=7.036\cdot10^{5}$ K. The rate of photoionization of atomic hydrogen from the first level by radiation with flux $F(\nu)$ we compute for the known effective cross-section of photoionization from the first level  $\sigma_1(\nu)$ according to \cite{Abel1997} as follows:
\begin{equation}
R_{1c}=\int_{\nu_{1c}}^\infty\sigma_1(\nu)\frac{F(\nu)}{h\nu}d\nu=c\int_{\nu_{1c}}^\infty\sigma_1(\nu)\frac{E_{\nu}}{h\nu}d\nu, \nonumber
\end{equation}
where $\nu_{1c}$ is the threshold frequency of photoionization from the first level which corresponds to the energy of ionisation of hydrogen atoms from the base state 13.6 eV.  We use the analytical approximation proposed by \cite{Verner1996a}, 
\begin{equation}
 \sigma_1(E)=\frac{\sigma_0(x-1)^2x^{P/2-5.5}}{\left(1+\sqrt{x/y_a}\right)^P},\nonumber
\end{equation}
where $\sigma_0=1.1083\cdot10^{-14}$ cm$^2$, $x=\frac{E}{E_0}$, $E=h\nu$, $E_0=1.0235$ eV,  $y_a=23.424$, $P=2.3745$ are the best-fit parameters taken from \cite{Novosyadlyj2022}.

We can estimate the number density of stars $N_s$ which are necessary to ionize all interstellar medium at $z=6$ using equation (\ref{ire}):
\begin{equation}
N_s\approx\frac{\alpha_r(T)x_{HII}^2n_Hn_*}{(1-x_{HII})R_{1c}},
\label{Ns_rei}
\end{equation}
where $n_*$ is number density of stars which we estimated in section I (multiplying by it, we subtract this value from $R_{1c}$) and $T$ is temperature of interstellar medium. We put it equals 10000 K (typical in Galaxy) and assume that all stars have appeared at $z=30$ and at $z=6$ $x_{HII}=0.99$. Integrating $R_{1c}$ over $\nu$ we obtain for three values of star temperatures 
\begin{eqnarray*}
N_s&\approx&4\cdot10^{16}\,\mbox{\rm kpc$^{-3}$}\quad \mbox{\rm for\,\,stars\,\,with}\quad T_*=5700\,\mbox{\rm K},\\ 
N_s&\approx&7\cdot10^{4}\,\mbox{\rm kpc$^{-3}$}\quad \mbox{\rm for\,\,stars\,\,with}\quad T_*=37500\,\mbox{\rm K}, \\
N_s&\approx&1\cdot10^{4}\,\mbox{\rm kpc$^{-3}$}\quad \mbox{\rm for\,\,stars\,\,with}\quad T_*=50000\,\mbox{\rm K}\\
\end{eqnarray*}
Our estimation of comoving number density of stars in section I is equal $n_*\sim700-7\cdot10^{4}$. On the way to this result, we made many simplifications and rough guesses, but we obtained a plausible result: solar-type stars cannot ensure complete ionisation of the interstellar medium at the end of the reionization epoch such as their number density should be implausibly large compared to the observed one. Conversely, young hot stars with a temperature of $\sim40000-50000$K (Population III stars) and a number density commensurate with the observed one could ionize the interstellar medium at $z=6$. Such stars have masses and radiuses  $\sim10-100$ or even more times large than the solar one, so, their number density would be $\sim10^2-10^4$ times lower than estimated above values (see expressions (\ref{Enu0})-(\ref{Enuz0}). 

\begin{figure}[htb!]
\includegraphics[width=0.48\textwidth]{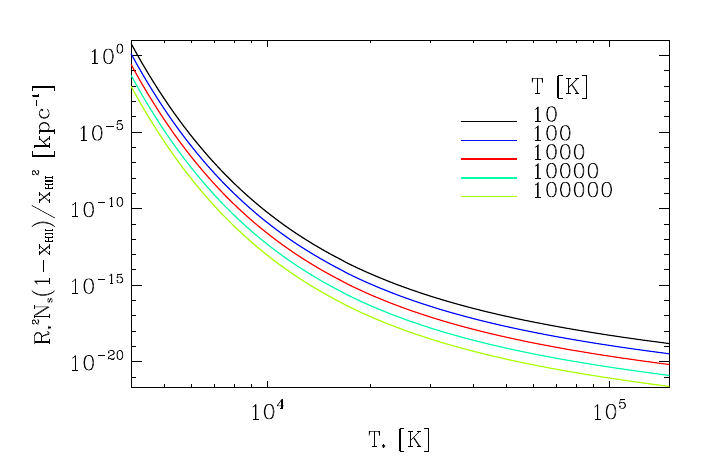}
\caption{The dependence of the product $R_*^2N_s(1-x_{HII})/x_{HII}^2$ on the effective temperature $T_*$ for a given value of temperature of interstellar medium $T=10,\,100,\,1000,\,10000,\,100000$ K (from top to down) in the  standard $\Lambda$CDM model.}
\label{fig8}
\end{figure}
The first billion year of life of the Universe are invisible for most Earth telescopes. We do not know which first sources ionized the intergalactic medium, which where their effective temperatures $T_*$, radius $R_*$ and number density $N_s$. If we multiply the left and right hand sides of equation (\ref{Ns_rei}) by the unknown value of the squared effective radius of the first light sources, $R_*^2$, and by the poorly determined by observations the ratio $(1-x_{HII})/x_{HII}$, then the right part of this equation will be only a function of the effective temperature of the sources $T_*$ and the temperature of the ionized medium $T$ at $z=6$. Repeating the computations of number density of sources with different $T_*$, which is necessary to ionize the medium at $z=6$, we obtain the dependence of the product $R_*^2N_s(1-x_{HII})/x_{HII}^2$ on the effective temperature $T_*$ for a given value of  temperature of medium $T$. Such dependence  for the temperature range $T_*=4000-150000$ K for a given value of temperature of interstellar medium $T=10,\,100,\,1000,\,10000,\,100000$ K is presented in fig. \ref{fig8}. It gives a possibility to estimate the number density of sources with any set ($R_*,\,T_*$), which is necessary to ionize the medium with some value $x_{HII}$ and $T$ at $z=6$. 

\begin{figure}[htb!]
\includegraphics[width=0.48\textwidth]{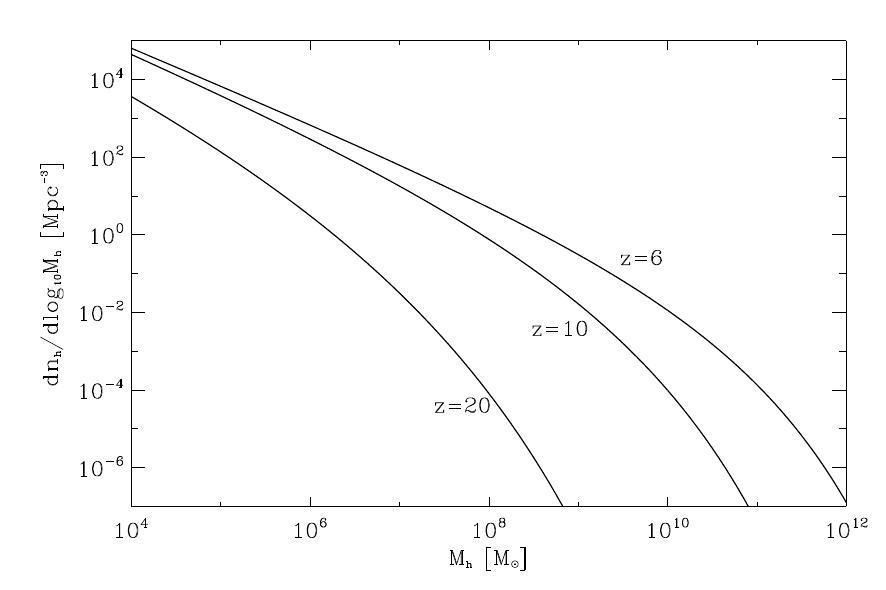}
\caption{The mass functions of halos virialised in Cosmic Dawn epoch in standard $\Lambda$CDM model at $z$=6, 10, 20.}
\label{fig9}
\end{figure}
Let's estimate the number density of sources at $z=6$ with some reasonable guesses about values of $R_*$, $T_*$ and $T$. Since, the baryonic medium in which the first stars were formed, was depleted of heavy elements, the masses of such stars must be large. Such stars are called the Population III stars (the first generation stars). Theorists assume that their mass may be 100 M$_\odot$ and more \cite{Bromm2011}. It is known that for massive stars of our galaxy with $M\ge10$ M$_\odot$ their masses, radiuses and temperatures are related as $R/R_\odot\approx2.67(M/M_\odot)^{0.36}$ and $T/T_\odot=(M/M_\odot)^{0.6}$. Assuming that they are correct for  Pop III stars we obtain the estimations for stars with 100 M$_\odot$: $R_*\approx14R_\odot\approx3.2\cdot10^{-10}$ kpc, $T\approx16T_\odot\approx86550$ K. The stars with such temperatures can heat the interstellar medium up to $T=10000$ K. Now we can estimate their number density using Fig. \ref{fig8}. The forth line from top for $T_*=86550$ K show that $R_*^2N_s(1-x_{HII})/x_{HII}^2\approx1.5\cdot10^{-20}$ kpc$^{-1}$. So, for estimated $R_*$ and $T_*$ we obtain $N_s(1-x_{HII})/x_{HII}^2\approx0.15$. Thus, the stars with $M=100M\odot$, $T=16T_\odot$, $R=14R\odot$ and number density one star per cubic kiloparsec can reionize the interstellar hydrogen at $z=6$ up to $x_{HII}\approx0.9$. The interstellar hydrogen at $z=6$ could be ionised up to $x_{HII}=0.99$ when the number density of such stars was $\approx$15 per cubic kiloparsec. 

This approach can be developed for other classes of luminous objects of the early Universe: globular star clusters and dwarf galaxies. For the globular clusters with radius 50 pc and effective temperature 10000 K we obtain $N_{gcl}\sim20$ Mpc$^{-3}$ and for the dwarf galaxies with radius 1 kpc and the same effective temperature $N_{dwg}\sim5\cdot10^{-2}$. These numbers do not disagree to initial power spectrum of density fluctuations deduced from the observational data on the cosmic microwave background radiation and large scale structure of the Universe. The mass function of virialised halos at $z$=6, 10 and 20 in the standard $\Lambda$CDM model is shown in Fig. \ref{fig9}. We computed them following the technic proposed in \cite{Schneider2012,Schneider2013}. One can see, that number densities of halos of mass of globular clusters ($M_{gcl}\sim10^6$ M$_\odot$) and dwarf galaxies ($M_{dwg}\sim10^8$ M$_\odot$) are essentially larger than our estimation. In the paper \cite{Atek2024} authors argue the dwarf galaxies as a main sources of UV radiation in the Cosmic Dawn and Reionization epochs.

\section*{CONCLUSIONS}

We have analysed the impact of the expansion of the cosmological redshift on the formation of total spectral energy density of radiation  in the expanding Universe filled by star-like sources with comoving number density $n_*$. Assuming that all sources have the blackbody spectrum with the same temperature $T_*$ we analysed the SEDR in some point, which is equidistant from the closest sources. 

We have obtained the general expressions for estimation of the averaged SEDR $E_{\nu}(z_o,z)$ at an arbitrary $z_o$ from the sources in spherical bulk $z> z_o$ as function of parameters of source luminocity and cosmological model. We have computed SEDR in the framework of the standard $\Lambda$CDM model with parameters from \cite{Planck2020} for sources with $T_*=5700$ and $T_*=50000$ for $z_o=0$ and $6$. 

The results show what the composite SEDR differ from the blackbody spectrum of sources which form it: the Wien part of the SEDR is decreased, the Rayleigh–Jeans one is increased with increasing the redshift of the edge of space volume filled by sources. This is due to the integration of radiation from sources located at different redshifts. The peak frequency of SEDR, $\nu_{peak}$, slowly decreases with increasing $(1+z)/(1+z_o)$, and practically freezes for $(1+z)/(1+z_o)>8$. The ratio of SEDR from stars, which are in the volume with comoving radius $r_i(z_i)$ in the expanding Universe, to SEDR from the same stars in the same volume in static infinite flat Universe, $E^{exp}_\nu/E^{stat}_\nu$, is $>1$ at lower frequencies, $\nu<0.3\div0.6\,\nu_{peak}$. It grows with increasing $z_i$.

The results for $z_o=0$ we have presented in the context of discussions about resolution of Olbers' paradox \cite{Harrison1964,Belinfante1975,Wesson1986,PanGen1988,Harrison1990,Wesson1991,Maddox1991,Peebles1993,Couture2012,Conselice2016,Mattila2019,Harari2019}. We therefore began with the analysis for the simple model of a homogeneous static infinite eternal universe for which this paradox arises. Our results prove that the Olbers` paradox in the standard Big Bang cosmological model is resolved due to 1) the finiteness of the observable volume of the expanding Universe (existing of a particle horizon), 2) the finiteness of the volume of the Universe filled with stars (radiation sources appeared in the epoch of Cosmic Dawn), and 3) the cosmological redshift.

Results for $z=6$ we present in the context of discussions about sources of reionization of intergalaxy gas at the end of the first billion years of our Universe  \cite{Bromm2011,Madau2015,Bouvens2015a,Robertson2015,Atek2015,Finkelstein2015,Bouvens2017,Dayal2018,Mason2019,Mistra2018,Ishigaki2018,Finkelstein2019,Naidu2020,Dayal2020,Robertson2022,Robertson2023,Bunker2023,Roberts2023,Mascia2023,Atek2024}. 
We have estimated the number density of stars, globular clusters and dwarf galaxies capable to provide the reionization of hydrogen atoms up to $x_{HII}\approx0.99$ at $z=6$. We show that a small part of number density of globular cluster and dwarf galaxies with thermal SEDR with temperature $T_*\sim10000$ K followed from the mass function of virialised halos of corresponding mass are capable to do that. 

\section*{Acknowledgements}
This work was supported by the International Center of Future Science and College of Physics of Jilin University (P.R.China), and the project of the National Research Found of Ukraine "Tomography of the Dark Ages and Cosmic Dawn in the lines of hydrogen and the firs molecules as a test of cosmological models” (state registration number 0124U004029).


\begin{thebibliography}{}
\bibitem {Bromm2011} V. Bromm, N. Yoshida, Rev. Astr. Astrophys., {\bf 49}, 373 (2011)  doi: 10.1146/annurev-astro-081710-102608
\bibitem {Madau2015} P. Madau, F. Haardt,  Astrophys. J. Lett. {\bf 813}, L8 (2015) doi: 10.1088/2041-8205/813/1/L8
\bibitem {Bouvens2015a} R. J. Bouwens et al., Astrophys. J. {\bf 803}, 34 (2015) doi: 10.1088/0004-637X/803/1/34
\bibitem {Robertson2015} B. E. Robertson, R. S. Ellis, S. R. Furlanetto and J. S. Dunlop, Astrophys. J. Lett. {\bf 802}, L19 (2015) doi: 10.1088/2041-8205/802/2/L19
\bibitem {Atek2015} H. Atek et al., Astrophys. J. {\bf814}, 69 (2015) doi: 10.1088/0004-637X/814/1/69
\bibitem {Finkelstein2015} S. L. Finkelstein et al.,  Astrophys. J. {\bf 810}, 71 (2015) doi: 10.1088/0004-637X/810/1/71
\bibitem {Bouvens2017} R. J. Bouwens, P. A. Oesch, G. D. Illingworth, R. S. Ellis and M. Stefanon,  Astrophys. J. {\bf 843}, 129 (2017) doi:  10.3847/1538-4357/aa70a4
\bibitem {Dayal2018} P. Dayal, A. Ferrara, Phys. Rep. {\bf 780–782}, 1 (2018) doi: 10.1016/j.physrep.2018.10.002
\bibitem {Mistra2018} S, Mitra, T. R. Choudhury and A. Ferrara, Mon. Not. R. Astron. Soc. {\bf 473}, 1416 (2018) doi: 10.1093/mnras/stx2443
\bibitem {Ishigaki2018} M. Ishigaki, R. Kawamata, M. Ouchi, M. Oguri, K. Shimasaku and Y. Ono, Astrophys. J. {\bf 854}, 73 (2018) doi: 10.3847/1538-4357/aaa544
\bibitem {Mason2019} C. A. Mason,R. P. Naidu, S. Tacchella and J. Leja, Mon. Not. R. Astron. Soc. {\bf 489}, 2669 (2019) doi: 10.1093/mnras/stz2291
\bibitem {Finkelstein2019} S. L. Finkelstein, A. D'Aloisio, J-P. Paardekooper, R. Ryan, P. Behroozi, K. Finlator, R. Livermore, P. R. Upton Sanderbeck, C. Dalla Vecchia and S. Khochfar,  Astrophys. J. {\bf 879}, 36 (2019) doi: 10.3847/1538-4357/ab1ea8
\bibitem {Naidu2020} R. P. Naidu, S. Tacchella, C. A. Mason, S. Bose, P. A. Oesch and C. Conroy, Astrophys. J. {\bf 892}, 109 (2020) doi: 10.3847/1538-4357/ab7cc9
\bibitem {Dayal2020} P. Dayal, M. Volonteri, T. R. Choudhury, R. Schneider, M. Trebitsch, N. Y. Gnedin, H. Atek, M. Hirschmann and A. Reines, Mon. Not. R. Astron. Soc. {\bf 495}, 3065 (2020) doi: 10.1093/mnras/staa1138
\bibitem {Robertson2022} B. E. Robertson, Annu. Rev. Astron. Astrophys. {\bf 60}, 121 (2022) 10.1146/annurev-astro-120221-044656
\bibitem {Robertson2023} B. E. Robertson et al., Nat. Astron. {\bf 7}, 611 (2023) doi: 10.1038/s41550-023-01921-1
\bibitem {Bunker2023} A. J. Bunker et al., arXiv:2306.02467 doi: 10.48550/arXiv.2306.02467
\bibitem {Roberts2023} G. Roberts-Borsani et al., Nature {\bf 618}, 480 (2023) doi: 10.1038/s41586-023-05994-w
\bibitem {Mascia2023} S. Mascia et al., Astron. Astrophys. {\bf 672}, A155 (2023) doi: 10.1051/0004-6361/202345866
\bibitem {Atek2024} H. Atek et al., Nature, {\bf 626}, 975 (2024) doi: 10.1038/s41586-024-07043-6
\bibitem {Harrison1964} E. R. Harrison, Nature, {\bf 204}, 271 (1964) doi: 10.1038/204271b0 
\bibitem {Belinfante1975} F. J. Belinfante, General Relativity and Gravitation {\bf 6}, 9 (1975) doi: 10.1007/BF00766594
\bibitem {Wesson1986} P.S. Wesson, Space Science Reviews, {\bf 44}, 169 (1986) doi: 10.1007/BF00227231
\bibitem {PanGen1988} G. Pan, Chinese Astron. Astrophys., {\bf 12}, 191 (1988) https://doi.org/10.1016/0275-1062(88)90046-X
\bibitem {Harrison1990} E.R. Harrison, \emph{The Dark Night-sky Riddle, ``Olbers' Paradox''}, in The Galactic and Extragalactic Background Radiation, Proceedings of the 139th. Symposium of the International Astronomical Union, held in Heidelberg, FRG, June 12-16, 1989. Eds. S. Bowyer, C. Leinert; Publishers, Kluwer Academic Publishers, Dordrecht, Boston, p.3 (1990)
\bibitem {Wesson1991} P. S. Wesson, Astrophys. J., {\bf 367}, 399 (1991) doi: 10.1086/169638
\bibitem {Maddox1991} J. Maddox, Nature, {\bf 349}, 363 (1991) doi: 10.1038/349363a0
\bibitem {Peebles1993} P. J. E. Peebles , Principles of Physical Cosmology, Princetone University Press, 718 p.(1993) doi: 10.1515/9780691206721
\bibitem {Couture2012} G. Couture, Eur. J. Phys., {\bf 33}, 479 (2012) doi: 10.1088/0143-0807/33/3/479
\bibitem {Conselice2016} C. J. Conselice, A. Wilkinson, K. Duncan and A. Mortlock, Astrophys. J., {\bf 830}, 83 (2016) doi: 10.3847/0004-637X/830/2/83
\bibitem {Mattila2019} K. Mattila, P. Väisänen, Contemporary Physics, {\bf 60}, 23 (2019) doi: 10.1080/00107514.2019.1586130
\bibitem {Harari2019} Z. Harari, Astronomische Nachrichten, {\bf 340}, 510 (2019) doi: 10.1002/asna.201913540
\bibitem {Planck2020} Planck Collaboration, Astron. Astrophys,  {\bf 641}, A1(2020) doi: 10.1051/0004-6361/201833880
\bibitem {Planck2020a}  Planck Collaboration , Astron. Astrophys,  {\bf 641}, A6 (2020) doi: 10.1051/0004-6361/201833910
\bibitem {Glazer2018} D. Glazer, M. M. Rau and H. Trac, Research Notes of the American Astronomical Society, {\bf 2}, 135 (2018) doi:  10.3847/2515-5172/aad68a
\bibitem {Schroeder2013} J. Schroeder, A. Mesinger and Z. Haiman, Mon. Not. R. Astron. Soc., {\bf 428}, 3058 (2013) doi: 10.1093/mnras/sts253
\bibitem {Greig2017} B. Greig, A. Mesinger, Z. Haiman and R. A. Simcoe, Mon. Not. R. Astron. Soc., {\bf 466}, 4239 (2017) doi: 10.1093/mnras/stw3351
\bibitem {Mortlock2011} D. J. Mortlock et al., Nature, {\bf 474}, 616 (2011)  doi: 10.1038/nature10159
\bibitem {Davies2018} F. B. Davies et al., Astrophys. J., {\bf 864}, 142 (2018) doi: 10.3847/1538-4357/aad6dc
\bibitem {Bouvens2015} R. J. Bouwens, G. D. Illingworth, P. A. Oesch, J. Caruana, B. Holwerda, R. Smit and S. Wilkins, Astrophys. J., {\bf 811}, 140 (2015) doi: 10.1088/0004-637X/811/2/140
\bibitem {Banados2018} E. Bañados et al., Nature, {\bf553}, 473 (2018) doi: 10.1038/nature25180
\bibitem {Greig2022} B. Greig, A. Mesinger, F. B. Davies, F. Wang, J. Yang and J. F. Hennawi,  Mon. Not. R. Astron. Soc., {\bf 512}, 5390 (2022) doi:10.1093/mnras/stac825
\bibitem {McGreer2015} I. D. McGreer, A. Mesinger and V. D'Odorico, Not. R. Astron. Soc., {\bf 447}, 499 (2015) doi: 10.1093/mnras/stu2449
\bibitem {Schenker2014} M. A. Schenker, R. S. Ellis R.S., N. P. Konidaris and D.P. Stark,  Astrophys. J., {\bf 795}, 20 (2014) doi: 10.1088/0004-637X/795/1/20
\bibitem {Mason2018} C. A. Mason, T. Treu, M. Dijkstra, A. Mesinger, M. Trenti, L. Pentericci, S. De Barros and E. Vanzella, Astrophys. J., {\bf 856}, 2 (2018) doi: 10.3847/1538-4357/aab0a7
\bibitem {Mason2019a} C. A. Mason et al., Mon. Not. R. Astron. Soc, {\bf 485}, 3947 (2019) doi: 10.1093/mnras/stz632
\bibitem {Ouchi2010} M. Ouchi et al., Astrophys. J., {\bf 723}, 869 (2010) doi: 10.1088/0004-637X/723/1/869
\bibitem {Gallerani2008} S. Gallerani, A. Ferrara, X. Fan and T. R. Choudhury, Mon. Not. R. Astron. Soc, {\bf 386}, 359 (2008) doi: 10.1111/j.1365-2966.2008.13029.x
\bibitem {Verner1996} D. A. Verner, G. J. Ferland, Astrophysical Journal, Supplement Series, {\bf 103}, 467(1996) doi: 10.1086/192284
\bibitem {Abel1997} T. Abel, P. Anninos, Y. Zhang, and M. L. Norman, New Astron., {\bf 2}, 181(1997) doi: 10.1016/S1384-1076(97)00010-9
\bibitem {Verner1996a} D. A. Verner, G. J. Ferland, K. Y. Korista and D. G. Yakovlev, Astrophys. J.,  {\bf 465}, 487(1996) doi: 10.1086/177435
\bibitem {Novosyadlyj2022} B. Novosyadlyj, Y. Kulinich, B. Melekh and V. Shulga, Astron. Astrophys, {\bf 663}, A120 (2022) doi: 10.1051/0004-6361/202243238
\bibitem {Novosyadlyj2023} B. Novosyadlyj, Y. Kulinich, G. Milinevsky and V. Shulga,  Mon. Not. R. Astron. Soc. {\bf 526}, 2724 (2023) doi:10.1093/mnras/stad2927
\bibitem {Novosyadlyj2024} B. Novosyadlyj, Yu. Kulinich, and O. Konovalenko, J. Phys. Stud. \textbf{28}, 1901 (2024) doi: https://doi.org/10.30970/jps.28.1901
\bibitem {Schneider2012} A. Schneider, R. E. Smith, A. V. Macciò and B. Moore, Mon. Not. R. Astron. Soc. {\bf 424}, 684 (2012) doi: 10.1111/j.1365-2966.2012.21252.x
\bibitem {Schneider2013} A. Schneider, R. E. Smith and D. Reed, Mon. Not. R. Astron. Soc. {\bf 433}, 1573 (2013) doi: 10.1093/mnras/stt829
\end{thebibliography}
\end{document}